\documentstyle[aps,preprint]{revtex}

\begin{document}

\setcounter{page}{0}

\title{The $d^*$ dibaryon in the extended quark-delocalization, 
color-screening model}

\author{Jialun Ping}
\address{Department of physics, Nanjing Normal University, Nanjing, 
210097, P.R. China;\\
Center for Theoretical Physics, Nanjing University, Nanjing, 210093, P.R.
China}
\author{Fan Wang}
\address{Center for Theoretical Physics and Department of Physics, 
Nanjing University, Nanjing, 210093, P. R. China}
\author{T. Goldman}
\address{Theoretical Division, Los Alamos National Laboratory, 
Los Alamos, NM 87545, USA}

\maketitle

\vspace{-5.0in}
\begin{flushright}{LA-UR-00-5688}\end{flushright}
\vspace{-0.40in}
\begin{flushright}{nucl-th/0012011}\end{flushright}
\vspace{4.0in}

\begin{abstract}
The quark-delocalization, color-screening model (QDCSM), extended by
inclusion of a one-pion-exchange (OPE) tail, is applied to the study of
the deuteron and the $d^*$ dibaryon. The results show that the
properties of the deuteron (an extended object) are well reproduced,
greatly improving the agreement with experimental data as compared to
our previous study (without OPE). At the same time, the mass and decay
width of the $d^*$ (a compact object) are, as expected, not altered
significantly.
\end{abstract}


\section{Introduction} 
Quantum Chromodynamics (QCD) is believed to be the fundamental theory
of the strong interactions. High energy phenomena can be described very
well by using its fundamental degrees of freedom: quarks and gluons.
However the direct use of QCD for low energy hadronic interactions, for
example the nucleon-nucleon ($NN$) interaction, is still impossible
because of the nonperturbative complications of QCD. Quark models are
therefore a useful phenomenological tool.  Which of the models or which
effective degrees of freedom represent the physics of QCD better
remains an open question.

The traditional meson-exchange
model~\cite{Machleidt1,Machleidt2,paris,HHG
} describes the $NN$
scattering data quantitatively very well, where the effective degrees
of freedom are nucleons and mesons. The intermediate- and long-range
parts of the $NN$ interaction are attributed to two-pion-exchange,
usually parameterized in terms of a $\sigma$ meson, and
one-pion-exchange (OPE).  The short-range part is either parameterized
by a repulsive core or regularized by means of vertex form factors.
Such parameterizations are difficult to extend to the study of new
phenomena, such as multiquark systems.

In light of its success in describing the properties of hadrons, the
constituent quark model (CQM)~\cite{CQM}, where the effective degrees
of freedom are constituent quarks and gluons, has been extended to the
study of the $NN$ interaction. The short-range repulsive core is
successfully reproduced by a combination of the quark Pauli exclusion
principle and the color hyperfine interaction.  On the other hand, the
intermediate- and long-range part of the $NN$ interaction cannot be
accounted for in the CQM. Meson-exchange has to be invoked again; this
leads to "hybrid" models~\cite{tubingen,tokyo,kyoto}. However, the
quantitative agreement with experimental data is not as good as for
traditional meson-exchange models.

Recently a new approach, the Goldstone-boson-exchange model~\cite{GBE},
where the effective degrees of freedom are constituent quarks and
Goldstone bosons, has appeared. It appears to give a rather good
description of the baryon spectrum and has also been applied to the
$NN$ interaction~\cite{Stancu}.

Another quark model approach, which is closer in spirit to the original
CQM, the quark-delocalization, color-screening model
(QDCSM)~\cite{prl}, has been developed with the aim of understanding
the well known similarities between nuclear and molecular forces
despite the obvious energy and length scale differences. The model has
been applied to baryon-baryon
interactions~\cite{prl,prc53,mpla10,npa657,npa673} and
dibaryons~\cite{prc51,mpla13,prc62}. Quantitative agreement with the
experimental data on $NN$ and $YN$ (hyperon) scattering has been
obtained~\cite{npa673}.

Although the intermediate range attraction of the $NN$ interaction is
reproduced in the model, the long range tail is missing, similar to the
case of the CQM.  For example, the deuteron, a highly extended object,
is not well reproduced ~\cite{dstar1}. An increased value for the
color screening parameter in the QDCSM can generate enough attraction
to bind the deuteron, but the radius and D-wave mixing thus obtained
are both too small.  Also, the attraction in $NN$ scattering is a
little bit too strong.  From these effects, from the parallel to the
CQM and from consideration of the coordinate space behavior of the
adiabatic $NN$ potential obtained in the model~\cite{npa673}, we
concluded that the long range part of the $NN$ interaction was what was
missing.  Additionally, the most convincing result of spontaneously
broken chiral symmetry is the small mass of the pion and its coupling
to the nucleon.  Pion exchange between nucleons is also uniquely well
established by partial wave analyses of $NN$ scattering data~\cite
{Timmermans}.

This paper reports a study of some effects of adding OPE to the QDCSM.
This is carried out with the inclusion of a short-distance coordinate
space cutoff in order to avoid, or at least, to minimize, double
counting of the intermediate to short range part of meson exchange
already accounted for in the model by delocalization of the quark wave
functions.  We first recalculate the deuteron, then apply this same
addition to the $d^*$ dibaryon. Another important application is that
to $NN$ scattering, but this is left to a future publication.  Sect.II
gives a brief description of the model Hamiltonian, wave functions and
calculation method. The results and a discussion are presented in
Sect.III.

\section{Model Hamiltonian, wave functions and calculation method} 
The details of the QDCSM can be found in Refs.\cite{prl,prc51} and the
resonating-group calculation method (RGM) has been presented in
Refs.\cite{dstar1,Buchmann}. Here we present only the model Hamiltonian,
wave functions and the necessary equations used in the current
calculation.

The Hamiltonian for the 3-quark system is the same as the usual potential
model. For the six-quark system, it is assumed to be
\begin{eqnarray} 
H_6 & = & \sum_{i=1}^6 (m_i+\frac{p_i^2}{2m_i})-T_{CM} +\sum_{i<j=1}^{6} 
    \left( V_{ij}^c + V_{ij}^G +V_{ij}^{\pi} \right) ,   \nonumber \\ 
V_{ij}^G & = & \alpha_s \frac{\vec{\lambda}_i \cdot \vec{\lambda}_j }{4} 
 \left[ \frac{1}{r_{ij}}-\frac{\pi \delta (\vec{r})}{m_i m_j} 
 \left( 1+\frac{2}{3} \vec{\sigma}_i \cdot \vec{\sigma}_j \right) 
  + \frac{1}{4m_im_j} \left( \frac{3(\vec{\sigma}_i \cdot 
 \vec{r}) (\vec{\sigma}_j \cdot \vec{r})}{r^5} - \frac{\vec{\sigma}_i \cdot 
  \vec{\sigma}_j}{r^3} \right) \right], \nonumber  \\
V_{ij}^{\pi} & = & \theta (r-r_0) f_{qq\pi}^2 \vec{\tau}_i \cdot \vec{\tau}_j
  \frac{1}{r} e^{-\mu_{\pi} r} \nonumber \\ 
  & & \hspace*{0.2in} \times \left[ \frac{1}{3} \vec{\sigma}_i \cdot 
  \vec{\sigma}_j + \left( \frac{3(\vec{\sigma}_i \cdot \vec{r}) 
  (\vec{\sigma}_j \cdot \vec{r})}{r^2} - \vec{\sigma}_i \cdot 
  \vec{\sigma}_j \right) \left( \frac{1}{(\mu_{\pi}r)^2} + 
  \frac{1}{\mu_{\pi}r} + \frac{1}{3} \right) \right], \label{hamiltonian} \\
V_{ij}^c & = & -a_c \vec{\lambda}_i \cdot \vec{\lambda}_j 
\left\{ \begin{array}{ll} 
 r_{ij}^2 & 
 \qquad \mbox{if }i,j\mbox{ occur in the same baryon orbit}, \\ 
 \frac{1 - e^{-\mu r_{ij}^2} }{\mu} & \qquad
 \mbox{if }i,j\mbox{ occur in different baryon orbits}, 
 \end{array} \right. \nonumber \\
\theta (r-r_0) & = & \left\{ 
 \begin{array}{ll}  0 & \qquad r < r_0, \\  1 & \qquad \mbox{otherwise}, 
 \end{array} \right. \nonumber 
\end{eqnarray} 
where all the symbols have their usual meaning, and the confinement
potential $V^C_{ij}$ has been discussed in Refs.\cite{prc62,dstar1}. 

%
The quark-pion coupling constant $f_{qq\pi}$ can be obtained from the
nucleon-pion coupling constant $f_{NN\pi}$ by using the equivalence of
the quark and nucleon pictures of the $NN$ interaction: If the
separation between two nucleons is large, then their interaction energy
can be well described by a Yukawa potential; the quark description of
the same separation should lead to the same potential. For example, for
the (IS)=(01) NN channel, (with the nucleon taken as a point particle,)
the Yukawa potential at separation R, is
\begin{equation}
V_{NN}^{\pi N} = - f_{NN\pi}^2 \frac{1}{R} e^{-\mu_{\pi} R}.
\end{equation}
Taking the nucleon as a $3q$ system, for sufficiently large separation,
$R$, we can express the potential between the two $3q$ systems as
\begin{eqnarray}
V_{NN}^{\pi q} & = & \langle [N(123)N(456)]^{IS} | 9 V_{14}^{\pi}
 |[N(123)N(456)]^{IS} \rangle  \nonumber \\
 & = &  - \frac{25}{9} f_{qq\pi}^2 \frac{1}{R} e^{-\mu_{\pi} R}
  e^{\mu_{\pi}^2 b^2/2} ,
\end{eqnarray}
since all of the exchange terms tend to zero at large $R$. The quantity
$b$ ($\sim 0.6$ fm) is the size parameter characterizing the quark wave
functions in a nucleon; see Refs.\cite{prc62,dstar1} and below. From
the above expression, it is clear that the classic symmetry relation,
$f_{qq\pi} = \frac{3}{5}f_{NN\pi}$, holds except for a small correction
(since $\mu_{\pi} b \sim 0.4$) due to the finite size of the nucleon in
the quark description.
%

After introducing generator coordinates to expand the relative motion
wave function and including the wave function for the center-of-mass
motion\footnote{For details, see Refs.\cite{dstar1,Buchmann}.}, the
ansatz for the two-cluster wave function used in the RGM can be written
as
\begin{eqnarray}
\Psi_{6q} & = & {\cal A} \sum_{k} \sum_{i=1}^n \sum_{L_k=0,2} C_{k,i,L_k} 
  \int \frac{d\Omega_{S_i}}{\sqrt{4\pi}}
  \prod_{\alpha=1}^{3} \psi_{\alpha} (\vec{S}_i , \epsilon) 
  \prod_{\beta=4}^{6} \psi_{\beta} (-\vec{S}_i , \epsilon)  \nonumber \\
  & & \left[ [\eta_{I_{1k}s_{1k}}(B_{1k})\eta_{I_{2k}s_{2k}}(B_{2k})]^{Is_k}
  Y^{L_k}(\hat{\vec{s}}_i) \right]^J [\chi_c(B_1)\chi_c(B_2)]^{[\sigma]}
	\label{multi} ,
\end{eqnarray}
where $k$ is the channel index. For example, for the deuteron, we have
$k=1, \ldots, 5$, corresponding to the channels $NN~ S=1~L=0$,
$\Delta\Delta~ S=1~L=0$, $\Delta\Delta~ S=3~L=2$, $NN~ S=1~L=2$, and
$\Delta\Delta~ S=1~L=2$. Also,
\begin{eqnarray}
\psi_{\alpha}(\vec{S}_i ,\epsilon) & = & \left( \phi_{\alpha}(\vec{S}_i) 
+ \epsilon \phi_{\alpha}(-\vec{S}_i)\right) /N(\epsilon), \nonumber \\
\psi_{\beta}(-\vec{S}_i ,\epsilon) & = & \left(\phi_{\beta}(-\vec{S}_i) 
+ \epsilon \phi_{\beta}(\vec{S}_i)\right) /N(\epsilon), \nonumber \\
N(\epsilon) & = & \sqrt{1+\epsilon^2+2\epsilon e^{-S_i^2/4b^2}}. \label{1q} \\
\phi_{\alpha}(\vec{S}_i) & = & \left( \frac{1}{\pi b^2} \right)^{3/4}
   e^{-\frac{1}{2b^2} (\vec{r}_{\alpha} - \vec{S}_i/2)^2} \nonumber \\
\phi_{\beta}(-\vec{S}_i) & = & \left( \frac{1}{\pi b^2} \right)^{3/4}
   e^{-\frac{1}{2b^2} (\vec{r}_{\beta} + \vec{S}_i/2)^2}. \nonumber
\end{eqnarray}
are the delocalized single-particle wave functions used in QDCSM. The
delocalization parameter, $\epsilon$, is determined by the six-quark
dynamics.

With the above ansatz, Eq.(\ref{multi}), the RGM equation
becomes an algebraic eigenvalue equation,
\begin{equation}
\sum_{j,k,L_k} C_{j,k,L_k} H^{k',L'_{k'},k,L_k}_{i,j} 
  = E \sum_{j} C_{j,k,L_k} N^{k',L'_{k'}}_{i,j}
   \label{GCM}
\end{equation}
where $N^{k',L'_{k'}}_{i,j}, H^{k,L_k,k',L'_{k'}}_{i,j}$ are the
(Eq.(\ref{multi})) wave function overlaps and Hamiltonian matrix
elements (without the summation over $L'$), respectively. By solving
the generalized eigen problem, we obtain the energies of the 6-quark
systems and their corresponding wave functions.

The partial width for $d^*$ decay into the $NN$ D-wave state is
obtained by using ``Fermi's Golden Rule'',
\begin{eqnarray}
\Gamma & = & \frac{1}{7} \sum_{M_{J_i}, M_{J_f}}
  \frac{1}{(2\pi)^2} \int p^2 dp~d\Omega~ \delta(E_f-E_i) |M|^2
  \nonumber \\
  & = & \frac{1}{7} \sum_{M_{J_i}, M_{J_f}}
  \frac{1}{32\pi^2} m_{d^*} \sqrt{m_{d^*}^2 - 4 m_N^2} \int |M|^2
  d\Omega,
  \label {width}
\end{eqnarray}
where $M_{J_i}$ and $M_{J_f}$ are the spin projections of the initial
and final states. The nonrelativistic transition matrix element, $M$,
includes the effect of the relative motion wave function between the
two final state nucleons,
\begin{equation}
M = \langle d^* | H_I | [\Psi_{N_1} \Psi_{N_2}]^{IS} e^{i \vec{p}\cdot \vec{R}}
 \rangle, \label{M}
\end{equation}
where $\vec{R}$ is the relative motion coordinate of the two clusters
of quarks (nucleons) and $p = \frac{1}{2}\sqrt{m_{d^*}^2 - 4m_N^2}$ is
the available relative momentum between the nucleons as determined by
the energy conserving $\delta$-function in Eq.(\ref{width}). The
interaction Hamiltonian, $H_I$, is comprised of the tensor parts of OGE
and OPE.

\section{Results and discussion}
Our model parameters are given in Table I.  They have been fixed by
matching baryon properties, except for the color screening parameter
($\mu$ in the confining potential in Eq.(\ref{hamiltonian})) which has
been determined by matching the mass of the deuteron. We have examined
the values 0.6 fm and 1.0 fm for the short-range cutoff of OPE.  For
each cutoff, the model parameters were readjusted to best match all of
the available data.  In all cases, the contribution of OPE to the
baryon mass is not large because of the short-range cutoff. Clearly
this is model dependent, as the mass of baryon comes mainly from OPE in
the GBE model~\cite{GBE}, and there is no net contribution from OPE in
the hybrid model of Fujiwara~\cite{kyoto}.

\begin{center}
Table I. Model parameters and calculated results for deuteron and $d^*$. 
\begin{tabular}{c|cc|cc|cc} \hline
 & \multicolumn{2}{c|}{$r_0=0.6$ fm} & \multicolumn{2}{c|}{$r_0=1.0$ fm} &
   \multicolumn{2}{c}{without OPE}  \\ \hline
 & deuteron & $d^*$ &  deuteron & $d^*$ & deuteron & $d^*$ \\ \hline
 $m$ (MeV) & \multicolumn{2}{c|}{313} & \multicolumn{2}{c|}{313} &
   \multicolumn{2}{c}{313}  \\
 $b$ (fm)& \multicolumn{2}{c|}{0.6010} & \multicolumn{2}{c|}{0.6021} &
   \multicolumn{2}{c}{0.6034}  \\ 
 $a_c$ (MeV fm$^{-2}$) & \multicolumn{2}{c|}{25.40} & 
   \multicolumn{2}{c|}{25.02} &
   \multicolumn{2}{c}{25.13}  \\
 $\alpha_s$  & \multicolumn{2}{c|}{1.573} & 
   \multicolumn{2}{c|}{1.550} &
   \multicolumn{2}{c}{1.543}  \\ \hline
 $\mu$ (fm$^{-2}$) & \multicolumn{2}{c|}{0.75} & 
   \multicolumn{2}{c|}{0.95} &
   \multicolumn{2}{c}{1.50}  \\ \hline
mass (MeV) &  1876 & 2186 &  1876 & 2165 &  1876 & 2116 \\
$\sqrt{\langle r^2 \rangle }$ (fm) & 2.1 & 1.3 & 1.9 & 1.3 & 1.5 & 1.2 \\
$P_D$ & 5.2\% & & 4.5\% & & 0.2\% &  \\
decay width (MeV) & & 7.92 & & 5.76 & & 4.02 \\ \hline
\end{tabular}
\end{center}

For comparison, the results of our earlier calculation~\cite{dstar1}
without OPE is also included in Table I.  Clearly, the ``deuteron''
obtained there is not the physically correct one, due to its small size
and negligible D-wave mixing, although it has the correct binding
energy. By adding OPE with a cutoff, our results are significantly
improved; the deuteron is now well reproduced with either cutoff scale.
Our calculations show that as the cutoff decreases, the deuteron size
and D-wave mixing both increase.  However, one cannot decrease the
cutoff to near zero as the D-wave mixing becomes too strong and the
size too large. This is presumably due to the double counting with the
quark delocalization effect already having accounted for the shorter
distance contributions of (more off-shell) pion exchange.  From the
overall good agreement, it seems that it may correct to conclude that
the quark delocalization and color screening mechanism also works well
in the short and intermediate ranges to replace the phenomenological
short-range repulsive core and $\sigma$-meson or double pion exchange
contributions of more conventional models.

However, the case of the $d^*$ is quite different. The OPE changes the
mass of the $d^*$ by only a few percent and increases the $NN$ decay
width by less than 4 MeV. These results were to be expected because of
the high degree of compactness of $d^*$, which can be seen from its
size, and the short-range cutoff of OPE.  The combination of these two
effects significantly reduces the contribution available from OPE to
the $d^*$.

In summary, we find that the deuteron can be well described in the
extended QDCSM. The quark delocalization and color screening mechanism
can account for the short-range repulsive core and most of the
intermediate-range attraction, while the missing long-range tail can be
economically incorporated by OPE with a short-range cutoff. With new
parameters thus fixed, the properties of the $d^*$ dibaryon are
minimally affected.

\vspace{0.2in}
This research is supported by the National Science Foundation of China,
the Fok Yingdung Educational Fund, the Natural Science Foundation of
Jiangsu Province and the U.S. Department of Energy under contract
W-7405-ENG-36.

\end{document}